# AI-Derived Reproductive Phenotypes and Explainable ML for Concurrent Early Multimorbidity in U.S. Women: NHANES 2017–March 2020

Sunday A. Adetunji, MD, MPH (Biostatistics)

PhD Student in Epidemiology/AI

College of Health, Oregon State University, Corvallis, Oregon, USA

# Abstract

Background: Adverse reproductive history is an emerging multisystemic risk factor; however, existing evidence is constrained by isolated outcome studies, limited covariate adjustment, and non-interpretable algorithmic models. This conference paper re-frames the estimand from future-event prediction to concurrent risk classification in a cross-sectional national sample and emphasizes calibration, interpretability, and systematic-error mitigation.

Methods: We analyzed 1,602 U.S. women aged 20–44 years from NHANES 2017–March 2020 with reproductive-history variables, chronic-condition indicators, and PHQ-9 data. The restricted early multimorbidity endpoint was defined as the presence of at least two of hypertension, hypercholesterolemia, cardiovascular disease, kidney disease, and kidney stones. Standardized reproductive, socioeconomic, chronic-disease, and depressive-symptom features were summarized using principal components analysis and grouped with *k*-means clustering. We compared multivariable logistic regression with gradient-boosted trees (XGBoost), evaluated discrimination and probability accuracy, and used SHAP values to quantify feature contributions.

Results: Early multimorbidity occurred in 6.6% (106/1,602) of the analytic cohort; 71.0% had no included chronic condition and 22.4% had one. Adverse reproductive burden was common, with 58% of women having at least one adverse reproductive factor and 12.6% having three or more;

mean PHQ-9 score was 3.7 (SD 4.4). Four latent phenotypes emerged (n = 398, 508, 102, and 594), including a clinically fragile subgroup in which 77.5% met the multimorbidity definition and mean PHQ-9 was 7.66. In-sample performance favored XGBoost over logistic regression (AUC 0.948 vs 0.742; Brier 0.038 vs 0.058). In holdout evaluation, XGBoost improved discrimination relative to logistic regression (ROC-AUC 0.766 [95% CI 0.683–0.848] vs 0.667 [0.562–0.772]; PR-AUC 0.217 vs 0.160), but showed worse probability accuracy and calibration (Brier 0.069 vs 0.059; log loss 0.280 vs 0.231; expected calibration error 0.113 vs 0.037). Across models, age, PHQ-9 score, income-to-poverty ratio, race/ethnicity, education, and the adverse reproductive index were dominant drivers.

Conclusions: Principal components analysis and k-means phenotyping revealed that adverse reproductive life-course structure is strongly clustered with concurrent early multimorbidity in U.S. women aged 20–44 years. Evaluation of XGBoost against logistic regression reveals that, while tree-based models may improve discrimination, superior AUC (Area Under the Receiver Operating Characteristic curve) does not equate to clinical, validated utility, highlighting that calibration and feature attribution are essential for reliable, credible translation into practice. This framework provides a rigorous basis for prospective temporal validation, high-validity endpoint construction, and clinically actionable multi-morbidity phenotypes to support future patient-level predictive risk stratification.

# 1. Introduction

Pregnancy is increasingly recognized as a natural physiological cardiovascular stress test, wherein adverse pregnancy outcomes (APOs) act as early sex-specific indicators for the unmasking, initiation, or accelerated progression of cardiovascular, cardiometabolic, renal, and systemic morbidities later in life, providing a critical window for risk stratification. Across large syntheses and cardiovascular statements, adverse pregnancy outcomes, pregnancy loss, gestational diabetes, hypertensive disorders of pregnancy, and related reproductive milestones have been linked to later cardiovascular disease, diabetes, chronic kidney disease, and psychosocial burden [10-15]. Yet many published analyses still rely on univariate outcomes, prioritize homogeneous or older-age cohorts, and frame machine learning as algorithmic ranking competitions rather than clinical decision-support systems (CDSS) requiring properly calibrated probabilities, interpretable feature attribution, and proper mitigation of confounding and dataset bias [1-9].

This knowledge gap is critical for women aged 20–44, a demographic often underrepresented in chronic disease surveillance due to low traditional morbidity prevalence. Yet, this period is characterized by cumulative reproductive and socioeconomic risks, and high depressive symptom burden, which collectively represent early-life predictors of future multisystem morbidity and valid targets for preventative stratification. A methodologically rigorous model for this setting should therefore do at least four things well: summarize heterogeneity without claiming to discover etiological subphenotypes, estimate probabilistic concurrent outcome risk with rigorous calibration assessment, explain individual-level predictions in clinically actionable language, and transparently acknowledge where cross-sectional secondary data constraints limit causal inference. [1-6,8,9].

This study develops and internally evaluates an interpretable, calibration-centered machine-learning framework for concurrent early multimorbidity classification in reproductive-age U.S. women. It treats reproductive history not as an isolated obstetric record, but as a structured life-course exposure architecture that may capture early multisystem vulnerability. Using nationally representative NHANES 2017–March 2020 data, we integrated principal components analysis, k-means phenotyping, multivariable logistic regression, gradient-boosted trees (XGBoost), SHAP-based attribution, and calibration-sensitive performance assessment to determine whether latent reproductive phenotypes concentrate early cardiometabolic, renal, and mental-health burden. By coupling unsupervised structure discovery with supervised classification and probability evaluation, the analysis moves beyond discrimination-only machine learning toward interpretable, epidemiologically grounded, and clinically credible risk stratification [1-8].

Our objectives were threefold: first, to derive clinically interpretable reproductive-health phenotypes in a national sample of U.S. women aged 20–44 years; second, to compare a parsimonious regression benchmark with a flexible boosted-tree classifier under a calibration-first evaluation framework; and third, to quantify the reproductive, socioeconomic, and mental-health features that most strongly drive the restricted early multimorbidity endpoint. The purpose was not to claim deployment readiness, but to establish an auditable analytic framework linking reproductive epidemiology, multimorbidity phenotyping, and explainable clinical machine learning, while providing a rigorous platform for external validation, broader endpoint construction, and subsequent longitudinal extension [1-15].

## 2. Methods

### 2.1 Study design, source population, and estimand

We conducted an analytic cross-sectional secondary analysis of pre-pandemic NHANES 2017–March 2020 public-use data. The study population was restricted to women aged 20–44 years with complete, non-missing information for reproductive-history, condition indicators, and the Patient Health Questionnaire (PHQ-9). The final analytic cohort comprised 1,602 women. Because the data are cross-sectional, the target estimand is not incident future risk, but rather the cross-sectional prevalence or the model-based concurrent probability that a woman meets the early multimorbidity definition at the time of assessment. Throughout this manuscript, terms such as *classification* and *stratification* are preferred over "*prospective prediction*" to reflect the analysis of concurrent data. [16,17] Reporting was guided by the TRIPOD+AI statement, adapted to the constraints of conference-length presentation. [1]

### 2.2 Restricted early multimorbidity endpoint

The primary endpoint was a restricted cardiorenal multimorbidity indicator defined as

$$Y_i = 1(M_i \geq 2), \qquad M_i = HYP_i + HCL_i + CVD_i + KID_i + KST_i.$$

Here $HYP$ denotes hypertension, $HCL$ hypercholesterolemia, $CVD$ cardiovascular disease, $KID$ kidney disease or dialysis, and $KST$ kidney stones. This definition employs a data-driven phenotypic classification to represent early multisystem burden. Given the intentional exclusion of diabetes, the endpoint is developed as a harmonized, composite construct (a 'proxy variable') designed specifically for the study's observational pipeline, representing a targeted sub-phenotype rather than an exhaustive mapping.

### 2.3 Predictor set and adverse reproductive index

Candidate covariates/predictors were selected to encompass cumulative reproductive history, socioeconomic position, and acute psychological distress. The core predictors were age, family income-to-poverty ratio, race/ethnicity, education, marital status, PHQ-9 total score, ever-pregnant status, gestational diabetes, pregnancy loss, parity-related information, and a composite adverse reproductive index. The index was defined as

$$A_i = GDM_i + MACRO_i + HPAR_i + LOSS_i + EXTAGE_i,$$

where $MACRO$ denotes macrosomic birth, $HPAR$ high parity, and $EXTAGE$ early or late age at first birth. By construction, larger values of $A_i$ represent greater clustering of adverse reproductive life-course events.

## 2.4 Latent phenotype derivation

To summarize heterogeneity without implying causality, predictors were standardized as

$$z_{ij} = \frac{x_{ij} - \bar{x}_j}{s_j}$$

and projected onto into a lower-dimensional representation with Principal Components Analysis (PCA). If $Z$ is the standardized design matrix, PCA solves the eigendecomposition (or SVD)

$$Sv_\ell = \lambda_\ell v_\ell, \qquad S = \frac{Z^\top Z}{n-1},$$

and participant scores on retained components are

$$u_{i\ell} = z_i^\top v_\ell.$$

$K$-means clustering was then applied to the retained score matrix $U$ by minimizing the within-cluster criterion

$$W(K) = \sum_{k=1}^{K} \sum_{i \in C_k} \|u_i - \mu_k\|^2.$$

A four-cluster $K = 4$ solution was selected, balancing interpretability with model parsimony. This partition should be interpreted as a descriptive typology rather than an objective ground truth. Optimal cluster partitioning was validated via a triangulation of internal metrics: assessing intra-cluster cohesion and inter-cluster separation (via average silhouette width), global partition structure (gap statistic), partition stability (non-parametric bootstrapping), and centroid consistency across random initializations. [19–21]

## 2.5 Classification models

The benchmark model was multivariable logistic regression:

$$\text{logit}\{\Pr(Y_i = 1 \mid X_i)\} = \beta_0 + X_i^\top \beta.$$

The flexible comparator was gradient-boosted trees, implemented as an additive ensemble in which the $t$-th update solves. [18]

$$\mathcal{L}^{(t)} = \sum_i \ell\left(y_i, \hat{y}_i^{(t-1)} + f_t(x_i)\right) + \Omega(f_t),$$

where $\ell(\cdot)$ is the Bernoulli deviance and $\Omega(f_t)$ penalizes tree complexity. We retained logistic regression not merely as a simplistic baseline, but as a clinically transparent, consistently well-calibrated reference model that remains competitive in rigorous, low-risk-of-bias clinical prediction settings [5-7].

## 2.6 Performance assessment

Discrimination was summarized with ROC-AUC and PR-AUC. Overall probability accuracy was quantified by the Brier score,

$$\text{Brier} = \frac{1}{n} \sum_{i=1}^{n} (y_i - \hat{p}_i)^2,$$

and the log loss

$$\text{LogLoss} = -\frac{1}{n} \sum_{i=1}^{n} [y_i \log(\hat{p}_i) + (1 - y_i)\log(1 - \hat{p}_i)].$$

Calibration was assessed using the calibration model

$$\text{logit}\{\Pr(Y_i = 1)\} = \alpha + \zeta \, \text{logit}(\hat{p}_i),$$

where $\alpha$ is the calibration intercept and $\zeta$ the calibration slope. Perfect calibration corresponds to $\alpha = 0$ and $\zeta = 1$. We also report the expected calibration error, defined over $M$ probability bins as

$$\text{ECE} = \sum_{m=1}^{M} \frac{|B_m|}{n} |\text{obs}(B_m) - \text{pred}(B_m)|.$$

Threshold-dependent clinical utility of the model was assessed using decision curve analysis (DCA) to measure net benefit using,

$$NB(t) = \frac{TP(t)}{n} - \frac{FP(t)}{n}\frac{t}{1-t},$$

where $NB(t)$ is the net benefit at threshold $t$ [5,8,12].

## 2.7 Explainability layer

For the Gradient-boosted tree model (GBDT), global feature importance and individual-level feature contributions were calculated with SHAP (SHapley Additive exPlanations) values. For an individual observation $x$, the model output was decomposed as

$$g(x) = \phi_0 + \sum_{j=1}^{p} \phi_j,$$

where $\phi_j$ is the Shapley contribution of feature $j$. Mean absolute SHAP values were used to rank global feature importance, while case-level contributions can be translated into interpretable patient-facing narratives without replacing statistical inference [9].

## 2.8 Bias structure and sensitivity framework

The principal threats to validity arise from outcome restriction, measurement error, missingness, overfitting, and transportability. First, because the endpoint includes only five conditions, the observed classification variable $Y^*$ is vulnerable to under-ascertainment relative to a broader latent burden state $Y$. Under a simple sensitivity model with near-perfect specificity and sensitivity $Se_Y$ for the observed endpoint, corrected prevalence would satisfy approximately

$$\Pr(Y = 1) \approx \frac{\Pr(Y^* = 1)}{Se_Y}.$$

Secondly, outcome ascertainment was treated as a measurement error problem rather than a purely administrative data-cleaning step. Since the restricted multimorbidity endpoint was derived from multiple self-reported items, incomplete responses were subjected to a probabilistic bias framework to quantify the potential for differential under-ascertainment of case status and subsequent attenuation of the association structure. Internal validation proceeded via a stratified development–holdout design, and comparative model assessment emphasized probability validity in addition to discrimination, using the Brier score, log loss, calibration intercept and slope, and expected calibration error alongside ROC-based metrics. We further included a bias map and quantitative sensitivity template to make the likely directions of systematic error analytically visible.

On this basis, the study presents an internally validated, calibration-optimized risk stratification model, adhering to transparent reporting standards, where model explainability, measurement fidelity, and systematic error auditing are integrated as core components of the inference framework [2,5,6].

## 3. Results

The analytic cohort included 1,602 women aged 20–44 years. Overall, 1,137 women (71.0%) had none of the included chronic conditions, 359 (22.4%) had one, and 106 (6.6%) met the restricted early multimorbidity definition. More than half of the cohort had at least one adverse reproductive factor, indicating that life-course reproductive burden clustered well before traditional older-age chronic-disease thresholds.

Four latent phenotypes were recovered from the PCA-plus-$k$-means pipeline and are summarized in Table 1 and illustrated conceptually in Figure 1. The K=4 solution showed the best balance of average silhouette width, gap statistic, and bootstrap stability, supporting its use as the descriptive phenotype partition for the present analysis The first phenotype represented socioeconomically constrained women with heavy reproductive burden but limited current multimorbidity, consistent with a pre-multimorbidity state. The second phenotype captured younger, socioeconomically advantaged women with the lowest overall burden. The third phenotype was distinctly clinically fragile, combining older age, concentrated adverse reproductive burden, depressive symptom elevation, and the greatest multimorbidity prevalence. The fourth phenotype reflected women in midlife with moderate reproductive burden, relatively low PHQ-9 scores, and little current multimorbidity.

The 'clinically fragile' phenotype manifested as the primary cluster of concurrent multisystem morbidity, evidenced by a mean age of 37.0 years, a mean PHQ-9 score of 7.66, and a prevalence of 77.5% (79/102) for the predefined multimorbidity endpoint. Conversely, the 'young advantaged' phenotype was characterized by a mean age of 26.8 years, a mean income-to-poverty ratio of 2.71, and a low cumulative incidence of the endpoint (11 of 508, 2.2%). These patterns suggest that reproductive history, structural position, and depressive symptom burden do not merely co-exist; they cluster into distinct empirical profiles associated with widely divergent contemporary disease burdens.[Table 1]

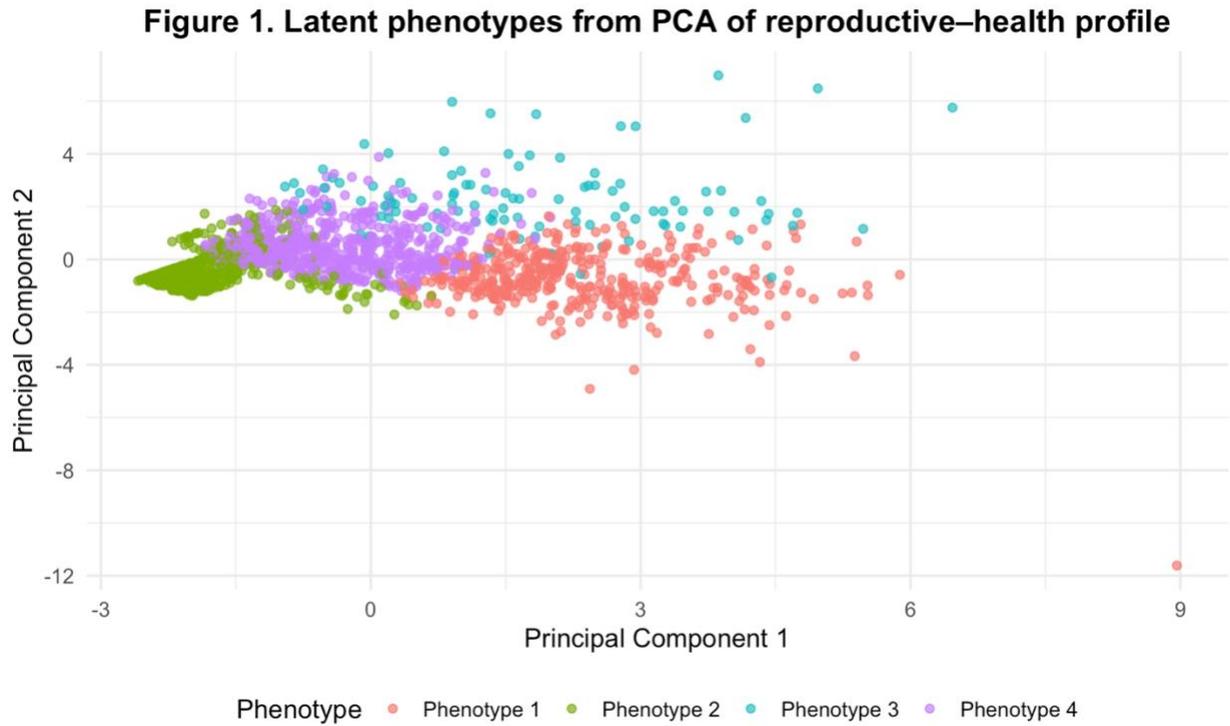

*Figure 1. Latent phenotypes of reproductive-chronic disease-mental-health profiles among U.S. women aged 20–44 years. Participants are projected onto the first two principal components of the standardized predictor matrix, and colors denote the four k-means-derived phenotypes. The purpose of the display is descriptive summarization of heterogeneity rather than causal subtype discovery.*

*Table 1. Phenotype summary for women aged 20–44 years in NHANES 2017–March 2020.*

| Phenotype label | n (%) | Age, y mean (SD) | PIR mean (SD) | PHQ-9 mean (SD) | >=2 conditions n (%) | Interpretation |
|---|---|---|---|---|---|---|
| P1: Constrained high-burden pre- | 398 (24 | 35.36 | 1.45 (1.13 | 3.85 | 7 (1.8) | Older, disadvantaged, high reproductive |

| Phenotype label | n (%) | Age, y mean (SD) | PIR mean (SD) | PHQ-9 mean (SD) | >=2 conditions n (%) | Interpretation |
|---|---|---|---|---|---|---|
| multimorbidity | .8) | (6.02) | ) | (4.56) | | burden with low current multimorbidity. |
| P2: Young advantaged low-burden | 508 (31.7) | 26.83 (5.65) | 2.71 (1.67) | 4.11 (4.71) | 11 (2.2) | Youngest and most socioeconomically advantaged profile. |
| P3: Clinically fragile multimorbid | 102 (6.4) | 37.00 (5.49) | 1.83 (1.26) | 7.66 (5.56) | 79 (77.5) | High reproductive burden, highest depressive symptoms, concentrated multimorbidity. |
| P4: Midlife moderate-burden low-depression | 594 (37.1) | 33.88 (6.58) | 2.69 (1.64) | 2.60 (3.12) | 9 (1.5) | Mid-aged, higher income, moderate reproductive burden, low depressive symptom load. |

*PIR denotes family income-to-poverty ratio. Phenotype labels are clinically interpretive names applied to the four k-means groups and do not imply causal disease subtypes.*

Model performance is summarized in Table 2 and visualized in Figure 2. In the development sample, XGBoost showed higher apparent performance than logistic regression (ROC-AUC 0.948 versus 0.742; Brier score 0.038 versus 0.058). However, the holdout comparison revealed a divergence in performance metrics across subgroups. XGBoost achieved better discrimination than logistic regression on the holdout sample (ROC-AUC 0.766 versus 0.667; PR-AUC 0.217 versus 0.160), but probability accuracy moved in the opposite direction. Logistic regression had the better Brier score (0.059 versus 0.069), lower log loss (0.231 versus 0.280), much better calibration intercept (-0.916 versus 1.588), better calibration slope (0.641 versus 2.954), and lower expected calibration error (0.037 versus 0.113).

Under a probabilistic calibration-first standard, this is the central empirical result of the study. The gradient-boosted tree model demonstrated superior discrimination, whereas the regression model provided superior calibration (closer agreement between predicted probabilities and observed event frequencies). For patient counseling, decision-threshold-based actions, and clinical deployment, this distinction is not cosmetic; it defines which model is more clinically appropriate for risk stratification.

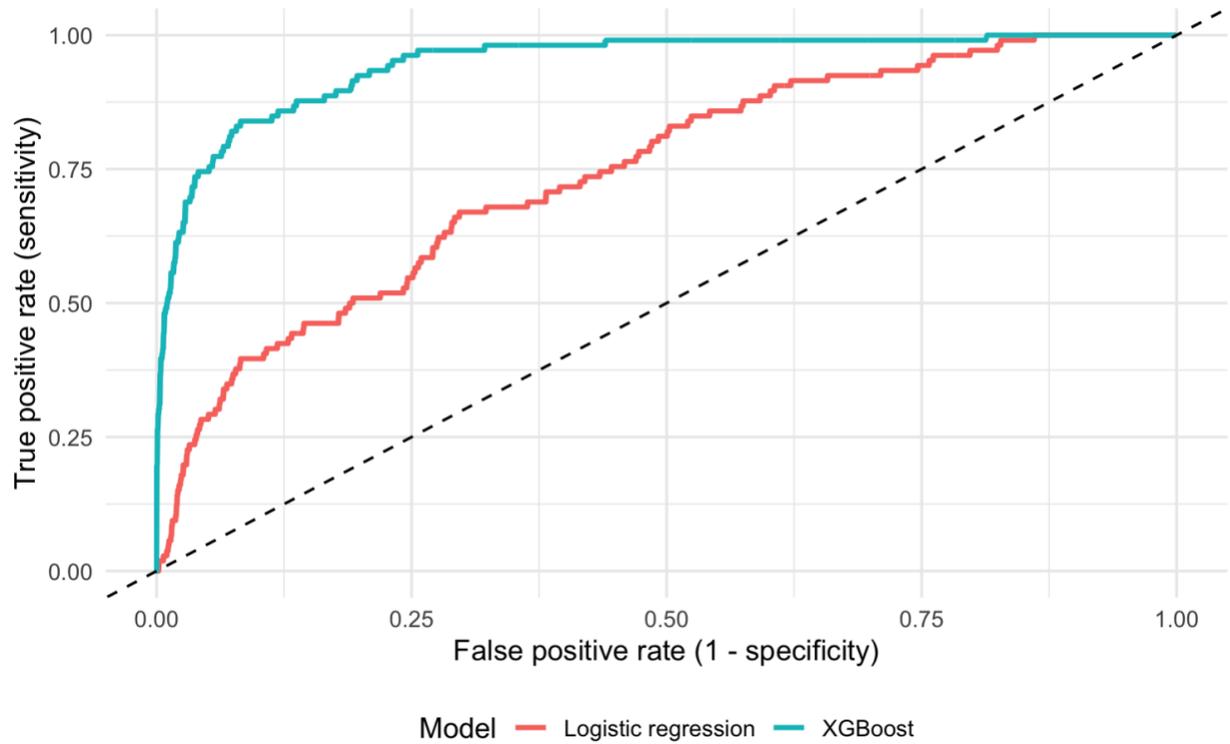

*Figure 2. Receiver-operating characteristic curves for restricted early multimorbidity classification comparing logistic regression with gradient-boosted trees (XGBoost). The diagonal line denotes chance performance. The key interpretation is not only the separation of curves but the mismatch between discrimination and calibration documented in Table 2.*

*Table 2. Development and holdout model performance.*

| Sample | Model | ROC-AUC | PR-AUC | Brier | Log loss | Calib. intercept | Calib. slope | ECE |
|---|---|---|---|---|---|---|---|---|
| Development | Logistic regression | 0.742 | — | 0.058 | — | — | — | — |
| Development | XGBoost | 0.948 | — | 0.038 | — | — | — | — |

| Sample | Model | ROC-AUC | | PR-AUC | Brier | Log loss | Calib. intercept | Calib. slope | ECE |
|---|---|---|---|---|---|---|---|---|---|
| Holdout | Logistic regression | 0.667 | (0.562–0.772) | 0.160 | 0.059 | 0.231 | -0.916 | 0.641 | 0.037 |
| Holdout | XGBoost | 0.766 | (0.683–0.848) | 0.217 | 0.069 | 0.280 | 1.588 | 2.954 | 0.113 |

*Development rows report the metrics available in the uploaded analysis summary. Full calibration metrics were available for the holdout comparison only. ECE denotes expected calibration error.*

Global explainability results are shown in Table 3 and Figure 3. The dominant predictors by mean absolute SHAP value were age (0.0407), PHQ-9 total score (0.0263), family income-to-poverty ratio (0.0255), race/ethnicity (0.0187), education (0.0117), and the adverse reproductive index (0.0076). Marital status, gestational diabetes, ever-pregnant status, and pregnancy loss contributed smaller but still non-trivial information. The ordering of this ranking is critical: the model did not demonstrate that chronic burden is driven solely by age or sociodemographic covariates. Instead, reproductive history and mental-health burden remained robust predictors within the model architecture.

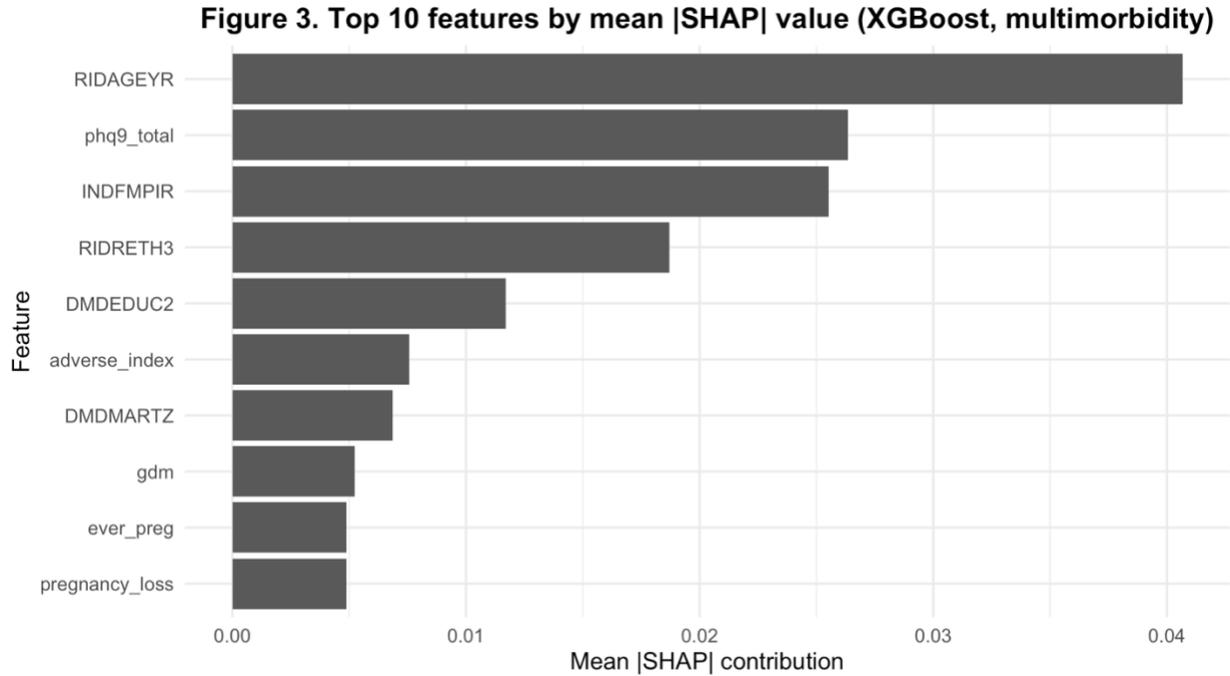

*Figure 3. Top ten features associated with restricted early multimorbidity in the XGBoost model. Bars represent mean absolute SHAP values and summarize average contribution magnitude across individuals. The plot conveys importance ranking, not the direction of effect for every individual.*

*Table 3. Top ten XGBoost predictors ranked by mean absolute SHAP value.*

| Rank | Feature | Mean |SHAP| |
|---|---|---|
| 1 | RIDAGEYR (age at screening) | 0.0407 |
| 2 | PHQ-9 total score | 0.0263 |
| 3 | INDFMPIR (income-to-poverty ratio) | 0.0255 |
| 4 | RIDRETH3 (race/ethnicity) | 0.0187 |
| 5 | DMDEDUC2 (education) | 0.0117 |
| 6 | Adverse reproductive index | 0.0076 |

| Rank | Feature | Mean \|SHAP\| |
|---|---|---|
| 7 | DMDMARTZ (marital status) | 0.0069 |
| 8 | History of gestational diabetes | 0.0052 |
| 9 | Ever pregnant | 0.0049 |
| 10 | Pregnancy loss | 0.0049 |

Decision-curve analysis (DCA) was undertaken to evaluate the clinical utility of predicted probabilities for informing decision thresholds. Across clinically plausible ranges, both models demonstrated superior net benefit compared to 'treat-all' and 'treat-none' strategies, indicating the estimated risks capture decision-analytic value not fully captured by standard discrimination measures (e.g., AUC). Consequently, DCA results support the framework's ability to guide targeted clinical management, complemented by calibration-sensitive evaluation and feature-attribution analysis.

## 4. Discussion

Three findings define the methodological and substantive contributions of this study. First, adverse reproductive history, socioeconomic position, and depressive symptom burden do not function as independent exposures; they aggregate into coherent phenotypic profiles, one of which concentrates a striking level of early multisystem comorbidity. Second, ensemble-based learning (gradient boosting) improved predictive discrimination (ranking) but not model calibration (probability validity), highlighting that a more flexible algorithm does not inherently improve clinical utility. Third, post-hoc explanation methods (SHAP) rendered the model's dependence on age, mental health, structural disadvantage, and reproductive burden mechanistically interpretable rather than merely probabilistic.

Empirical validation of calibration should take precedence in model evaluation. Modern predictive modeling guidance emphasizes that models intended for clinical decision support (CDS) or threshold-based risk stratification must produce absolute risk probabilities that are clinically actionable (well-calibrated), rather than merely relying on discrimination (rank-ordering) performance. [1-8,11,12]. In the present analysis, the XGBoost model demonstrated superior discriminative performance (AUC-ROC and AUC-PR) on the holdout sample, yet logistic regression maintained superior probabilistic calibration, as evidenced by a lower Brier score and stronger calibration slope and lower log loss. This highlights the clinical limitation of relying exclusively on rank-based metrics for model selection: a model with higher overall discrimination can be inferior for informing individual patient decisions due to poor calibration.

The phenotype or subtype structure reveals profound epidemiologic significance. The identified clinically vulnerable phenotype is characterized by advanced age, moderate socioeconomic disadvantage, accumulated reproductive burden, and the highest PHQ-9 scores. This aligns with literature identifying adverse reproductive history as an early prognostic biomarker for cardiometabolic and renal dysregulation. Furthermore, these findings suggest that reproductive and mental health burdens show convergence well before conventional clinical surveillance intensifies [10-15]. Rather than establishing direct causal inference, this study provides a high-dimensional, unsupervised phenotypic mapping of population-level data, characterizing distinct subphenotypes and their stochastic co-occurrence patterns.

The limitations are equally important. The inferential scope of this analysis is shaped by the endpoint definition, measurement structure, and validation design. We intentionally defined early multimorbidity as a restricted composite of hypertension, hypercholesterolemia, cardiovascular disease, kidney disease, and kidney stones, rather than as a fully comprehensive chronic-disease endpoint. Diabetes mellitus was not included in that composite because gestational diabetes history was already represented in the predictor space as a reproductive exposure and as a component of the adverse reproductive index; simultaneous inclusion of closely related diabetes constructs in both predictor and outcome domains would weaken exposure-outcome separation and create the possibility of target leakage, mechanically amplified association structure, and overstated predictive performance. The endpoint components were self-reported and thus vulnerable to misclassification, and the current handling of incomplete condition responses may permit some degree of under-ascertainment.

Additional considerations follow from the data source and validation design. Endpoint components were derived from self-reported condition indicators and are therefore subject to measurement error and incomplete ascertainment. Missing condition responses were handled within the current analytic pipeline in a way that may favor under-ascertainment of endpoint prevalence under some missingness structures. The phenotype architecture was derived from principal components analysis and k-means clustering without full perturbation-based stability characterization, and internal validation relied on a single stratified development-holdout split rather than repeated resampling. Finally, because NHANES 2017–March 2020 is cross-sectional, the resulting probabilities should be interpreted as concurrent classification estimates rather than prospective incidence predictions.

Accordingly, this analysis is best understood as an internal validation, calibration-optimized risk stratification model study. Natural next steps include rigorous bootstrap-based internal validation, formal cluster-stability and cluster-number (K) selection diagnostics, explicit model updating (recalibration) procedures, robust missingness analyses, and temporal/external sensitivity analyses that examine how the addition of diabetes mellitus changes model performance, prevalence, variable importance, and risk stratification. Using longitudinal, temporally ordered data, the framework can be extended from static risk stratification to dynamic prognostic modeling, featuring external validation and transportability assessment.

# 5. Conclusion

In U.S. women aged 20–44 years, cumulative reproductive life-course factors, structural disadvantage, and depressive symptom severity act as combined determinants that synergistically predict early-onset multimorbidity. Although the boosted-tree model yielded superior discrimination, the benchmark regression model maintained better calibration, reinforcing that clinical deployment requires prioritizing well-calibrated probabilistic predictions over higher rank-order performance alone. This evidence supports a probabilistically calibrated, bias-mitigated, and interpretable classification architecture, establishing a rigorous methodological foundation while leaving external validation, generalizability assessment, and implementation research to future study

# Ethics approval, participant protections, and data availability

This study was a secondary analysis of the public-use, de-identified NHANES 2017–March 2020 pre-pandemic files. NHANES protocols undergo review by the NCHS Ethics Review Board, and the survey uses documented informed consent procedures; CDC also provides cycle-specific consent materials and participant documents for the 2017–March 2020 release. The analytic dataset used here comes from the CDC/NCHS 2017–March 2020 pre-pandemic release, which combines 2017–2018 data with 2019–March 2020 data to support nationally representative pre-pandemic estimation, together with publicly posted analytic guidance for weighting and design-based inference. No new participant recruitment, contact, or intervention was undertaken for this secondary analysis. Data files, documentation, consent materials, and analytic guidance are publicly accessible from CDC/NCHS. *https://wwwn.cdc.gov/nchs/nhanes/continuousnhanes/default.aspx?Cycle=2017-2020*

# References

1. Collins GS, Moons KGM, Dhiman P, Riley RD, Beam AL, Van Calster B, et al. TRIPOD+AI statement: updated guidance for reporting clinical prediction models that use regression or machine learning methods. *BMJ. 2024;385:q902. doi:10.1136/bmj.q902*

2. Moons KGM, Damen JAA, Kaul T, Hooft L, Andaur Navarro C, Dhiman P, et al. PROBAST+AI: an updated quality, risk of bias, and applicability assessment tool for prediction models using regression or artificial intelligence methods. *BMJ*. 2025;388:e082505. doi:10.1136/bmj-2024-082505.

3. Collins GS, Dhiman P, Ma J, Schlussel MM, Archer L, Van Calster B, et al. Evaluation of clinical prediction models (part 1): from development to external validation. *BMJ*. 2024;384:e074819. doi:10.1136/bmj-2023-074819.

4. Efthimiou O, Seo M, Chalkou K, Debray TPA, Egger M, Salanti G. Developing clinical prediction models: a step-by-step guide. *BMJ*. 2024;386:e078276. doi:10.1136/bmj-2023-078276.

5. Riley RD, Collins GS, Kirton L, Snell KIE, Ensor J, Whittle R, et al. Uncertainty of risk estimates from clinical prediction models: rationale, challenges, and approaches. *BMJ*. 2025;388:e080749. doi:10.1136/bmj-2024-080749.

6. Van Calster B, McLernon DJ, Van Smeden M, Wynants L, Steyerberg EW, Bossuyt P, et al. Calibration: the Achilles heel of predictive analytics. *BMC Med*. 2019;17(1):230. doi:10.1186/s12916-019-1466-7.

7. Christodoulou E, Ma J, Collins GS, Steyerberg EW, Verbakel JY, Van Calster B. A systematic review shows no performance benefit of machine learning over logistic regression

for clinical prediction models. *J Clin Epidemiol.* 2019;110:12-22. doi:10.1016/j.jclinepi.2019.02.004.

8. Van Calster B, Wynants L, Verbeek JFM, Verbakel JY, Christodoulou E, Vickers AJ, et al. Reporting and interpreting decision curve analysis: a guide for investigators. *Eur Urol.* 2018;74(6):796-804. doi:10.1016/j.eururo.2018.08.038.

9. Lundberg SM, Erion G, Chen H, DeGrave A, Prutkin JM, Nair B, et al. From local explanations to global understanding with explainable AI for trees. *Nat Mach Intell.* 2020;2:56-67. doi:10.1038/s42256-019-0138-9.

10. Okoth K, Chandan JS, Marshall T, Thangaratinam S, Thomas GN, Nirantharakumar K, Adderley NJ. Association between the reproductive health of young women and cardiovascular disease in later life: umbrella review. *BMJ.* 2020;371:m3502. doi:10.1136/bmj.m3502.

11. Parikh NI, Gonzalez JM, Anderson CAM, Judd SE, Rexrode KM, Hlatky MA, et al. Adverse pregnancy outcomes and cardiovascular disease risk: unique opportunities for cardiovascular disease prevention in women: a scientific statement from the American Heart Association. *Circulation.* 2021;143(18):e902-e916. doi:10.1161/CIR.0000000000000961.

12. Garovic VD, White WM, Vaughan L, Saiki M, Parashuram S, Garcia-Valencia O, et al. Incidence and long-term outcomes of hypertensive disorders of pregnancy. *J Am Coll Cardiol.* 2020;75(18):2323-2334. doi:10.1016/j.jacc.2020.03.028.

13. O'Kelly AC, Michos ED, Shufelt CL, Vermunt JV, Minissian MB, Quesada O, et al. Pregnancy and reproductive risk factors for cardiovascular disease in women. *Circ Res.* 2022;130(4):652-672. doi:10.1161/CIRCRESAHA.121.319895.

14. McNestry C, Killeen SL, Crowley RK, McAuliffe FM. Pregnancy complications and later life women's health. *Acta Obstet Gynecol Scand.* 2023;102(5):523-531. doi:10.1111/aogs.14523.

15. Quenby S, Gallos ID, Dhillon-Smith RK, Podesek M, Stephenson MD, Fisher J, et al. Miscarriage matters: the epidemiological, physical, psychological and economic burden of early pregnancy loss. *Lancet.* 2021;397(10285):1658-1667. doi:10.1016/S0140-6736(21)00682-6.

16. Stierman B, Afful J, Carroll MD, Chen TC, Davy O, Fink S, et al. National Health and Nutrition Examination Survey 2017–March 2020 prepandemic data files—development of files and prevalence estimates for selected health outcomes. Natl Health Stat Report. 2021;(158). doi:10.15620/cdc:106273.

17. Akinbami LJ, Chen TC, Davy O, Ogden CL, Fink S, Clark J, et al. National Health and Nutrition Examination Survey, 2017–March 2020 prepandemic file: sample design, estimation, and analytic guidelines. Vital Health Stat 2. 2022;(190):1-36. doi:10.15620/cdc:115434.

18. Chen T, Guestrin C. XGBoost: a scalable tree boosting system. In: Proceedings of the 22nd ACM SIGKDD International Conference on Knowledge Discovery and Data Mining. New York, NY: ACM; 2016:785-794. doi:10.1145/2939672.2939785.

19. Rousseeuw PJ. Silhouettes: a graphical aid to the interpretation and validation of cluster analysis. J Comput Appl Math. 1987;20:53-65. doi:10.1016/0377-0427(87)90125-7.

20. Tibshirani R, Walther G, Hastie T. Estimating the number of clusters in a data set via the gap statistic. J R Stat Soc Series B Stat Methodol. 2001;63(2):411-423. doi:10.1111/1467-9868.00293.

21. Hennig C. Cluster-wise assessment of cluster stability. Comput Stat Data Anal. 2007;52(1):258-271. doi:10.1016/j.csda.2006.11.025.